\documentclass[12pt]{article}
\usepackage{amsmath,amsthm,amsfonts,amssymb, graphicx,epsfig}

\usepackage[usenames]{color}
\usepackage[active]{srcltx}

\newcommand{\ket}[1]{\left| #1 \right\rangle}
\newcommand{\bra}[1]{\left\langle #1 \right|}
\newcommand{\braket}[2]{\langle #1 | #2 \rangle}

\newcommand{\pd}{\partial_}

\newcommand{\re}{\operatorname{\mathrm{Re}}}
\newcommand{\im}{\operatorname{\mathrm{Im}}}
\newcommand{\Ran}{\operatorname{\mathrm{Range}}}
\newcommand{\Ker}{\operatorname{\mathrm{Ker}}}
\newcommand{\Gr}{\operatorname{\mathrm{Gr}}}
\newcommand{\rank}{\operatorname{\mathrm{rank}}}

\def\tr{\mathrm{Tr}}
\newcommand{\e}{\varepsilon}
\newcommand{\dbar}{\kern-.1em{\raise.8ex\hbox{ -}}\kern-.6em{d}}
  
\def\half{\mbox{$\frac 1 2$}}

\def\lind{{\cal L}}

\def\?{\marginpar{not sure}}
\newtheorem{thm}{Theorem}

\newtheorem{prop}{Proposition}

\def \be{\begin{equation}}
\def \ee{\end{equation}}
\def \bea{\begin{eqnarray}}
\def \eea{\end{eqnarray}}

\begin{document}



\title{Quantum response of dephasing open systems}
\author{J.E.~Avron${}^1$, M.~Fraas${}^1$, G.M.~Graf${}^2$ and O.~Kenneth${}^{1,3}$\\
{\small ${}^1$ Department of Physics, Technion, 32000 Haifa, Israel}\\
{ \small ${}^2$ Theoretische Physik, ETH Zurich, CH--8093 Z\"urich}  \\
{ \small ${}^3$ School of Physics and Astronomy, Tel-Aviv University, Tel Aviv 69978, Israel}
}




\maketitle
\begin{abstract}
We develop a  theory of adiabatic response for open systems governed by
Lindblad evolutions. The theory determines the dependence of the response
coefficients on the dephasing rates and allows for residual dissipation {\em
even when the ground state is protected by  a spectral  gap}. We give quantum
response a geometric interpretation in terms of Hilbert space projections: For
a two level system and, more generally, for systems with suitable functional form of the dephasing, the dissipative and non-dissipative parts of the response are linked to a metric and to a symplectic form. The metric is the Fubini-Study metric and the symplectic form is the adiabatic curvature.  When the metric and symplectic structures are {\em compatible}  the non-dissipative  part of the inverse matrix of response coefficients  turns out to be {\em immune to dephasing}.  We give three examples of physical systems whose quantum states induce compatible metric and symplectic structures on control space: The qubit, coherent states and a model of the integer quantum Hall effect.
\end{abstract}

Two frameworks that provide insight and understanding of transport coefficients are  Kubo's theory of linear response \cite{Mahan,EvansMorriss} and the theory of adiabatic response \cite{thbook,BR93,AvronRavehZur}. Both have a quantum version and a classical version, agree when there is overlap, and endow the {\em non-dissipative} transport coefficients with the geometric meaning of the adiabatic (Berry's) curvature \cite{berry}.

A notable success of this approach to ``geometrization of quantum response''  has been its application  to the integer quantum Hall effect where one observes  quantized resistances of the form $h/ne^2$ with $n$ an integer. The accuracy and robustness of the quantization is understood as reflecting the nontrivial topology of the quantum state characterized by a topological invariant, the Chern number, which is the $n$ measured as the Hall resistance \cite{tknn}.
 ``Geometrization of transport'' also lies at the heart of topological classification of states of matter \cite{AvronSeilerSimon,Zirnbauer,Kitaev} and the considerable current interest in topological insulators \cite{KaneMele}.

Our aim here is to  carry over the program of ``geometrization of quantum response'' to open quantum systems.  More precisely, extend the theory of adiabatic quantum response from the Hamiltonian setting to the Lindbladian setting \cite{Davies,BreuerPetruccione}.
The extension  gives geometric meaning to both the non-dissipative and dissipative response coefficients, 
and allows  to examine how they are affected by decoherence. It allows to address the stability of topological quantum numbers, such as Chern numbers, against dephasing.



\begin{figure}[htb]
    \centering
  \includegraphics[height=4 cm]{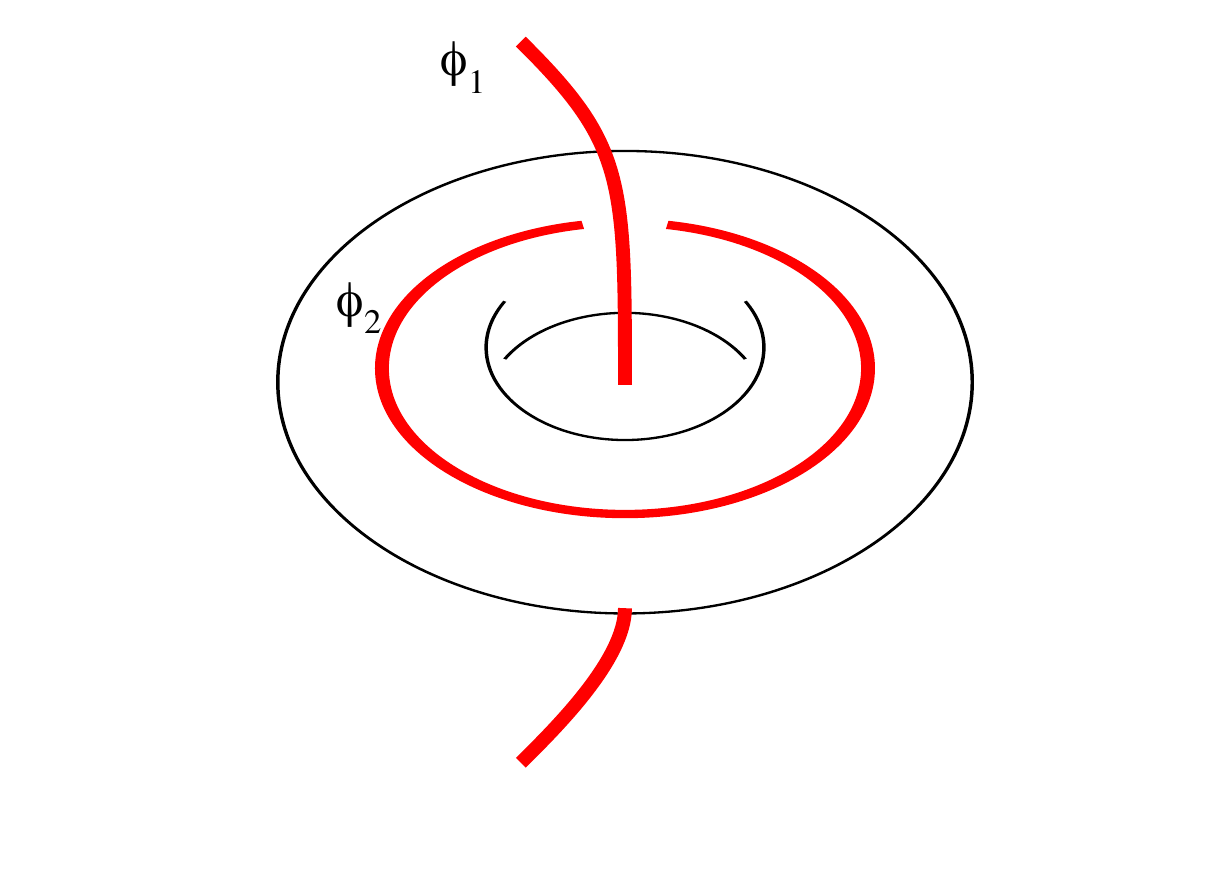}
    \caption{ A torus in coordinate space threaded by two loops (red) carrying fluxes $\phi_1$ and $\phi_2$ viewed as a model for the QHE. The (strong) magnetic field (not shown) is perpendicular to the surface of the torus. The control space ${\cal M}$ is the space of fluxes. By the Aharonov-Bohm periodicity, ${\cal M}$ is a torus as well. }\label{fig:torus}
    \end{figure}

In adiabatic response,  the Hamiltonian, $H(\phi)$, is viewed as a function of control parameters $\phi$ which drive  the system \cite{thbook,BR93,AvSeYa}. The space of controls shall be denoted by ${\cal M}$.  We focus on observables which are gradients of $H$ over ${\cal M}$, namely
	\be\label{Fmu}
	 F_\mu=\frac{\partial H}{\partial{\phi^\mu}}.
	\ee
For example,  in the system shown in Fig.~\ref {fig:torus}, the controls are magnetic fluxes which by the Aharonov-Bohm effect may be thought of as angular variables so that control space ${\cal M} $ is the torus $\mathbb {T}^2$.
$F_\mu$ is the loop-current in the $\mu$ loop as one can see by the principle
of virtual work. For other notions of currents in the Lindbladian context, see~\cite{Bel}.

Our main object of interest is  the matrix of response coefficients $f$  defined through the {\em linear  and instantaneous} terms in $\dot\phi$ in the (adiabatic) expansion of the response
	\be\label{LinResp}
	\tr\left(\rho_t F_\mu\right)=
	 f_{\mu\nu}\, \dot \phi^\nu(t) + \dots\,.
	\ee
Summation convention over repeated indices is implied. All other terms, which are not proportional to $\dot \phi$, do not
concern us.  In the example of Fig.~\ref {fig:torus}
$\dot\phi^\mu$ is  the emf on the $\nu$ loop and  $f$ is then the conductance matrix relating loop currents to emf's.

The quantum state $\rho_t$ in Eq. (\ref{LinResp}) arises from the ground state
as the solution at time $t$ of the equation of motion
$$
 	  \frac {d \rho}{dt} =\lind (\rho)	
$$
governed by the adiabatically changing {\em Lindbladian}:
\be\label{Lind}
\lind (\rho) = -i[H,\rho] +
\sum_{\alpha}\bigl([\Gamma_\alpha\rho,\Gamma_\alpha^*]+
[\Gamma_\alpha,\rho\Gamma_\alpha^*]\bigr);
	\ee
$\Gamma_\alpha=0$ corresponds to a unitary evolution.

In adiabatic transport \cite{thbook,BR93,AvronRavehZur} one is interested in the situation where the  control parameters $\phi$ move adiabatically along a path in control space  $ {\cal M}$. It turns out that adiabatic methods used to study the unitary case, $\Gamma_\alpha =0$,  can be extended to also study open systems described by Lindbladians, provided the instantaneous stationary states move continuously with the controls. The response of a stationary state, $\lind(\rho(\phi))=0$, of the Lindbladian $\lind$ to a driving is given by (see Appendix~B)
\begin{equation}
\label{fmunu}
f_{\mu \nu} = \tr\left(F_\mu \lind^{-1}(\pd{\nu}\rho)\right).
\end{equation}

$f_{\mu \nu}$ has a  geometric interpretation when the stationary
state $\rho$ is a (spectral) projection. A particular choice of Lindbladians that
achieve this is  $\Gamma_\alpha =\Gamma_\alpha (H)$  for some function of $H$.
We call this  family of changing Lindbladians, where  $\Gamma_\alpha (H)$ are slaved to the {\em instantaneous} Hamiltonian, {\em dephasing Lindbladians}.
The family describes processes that conserve energy and
 entropy increases without heat exchange.
  In appendix  A we give an example showing how  dephasing Lindbladians naturally emerge in certain stochastic unitary evolutions.



The slaving of the decoherence terms to the instantaneous Hamiltonian has the consequence that the instantaneous stationary states  of the Hamiltonian and the Lindbladian coincide. Moreover, when the controls $\phi$ vary adiabatically, one expects spectral projections of $H$   to evolve so that they remain close to the corresponding instantaneous spectral projections $P(\phi)$. The evolution of spectral projections can be described in geometric terms.
Geometric quantum response is concerned with the relation  between the response matrix $f$ and the geometry of $P(\phi)$ on control space.

Before we describe this relation, it is convenient to review the geometry on the control space
associated with a single spectral bundle $\phi\mapsto P(\phi)$.
Consider the (operator-valued) 1-form giving the natural adiabatic connection (here $P_\perp = 1-P$)
	\be\label{eq:A}
	A=A_\mu \, d\phi^\mu,\quad A_\mu=P_\perp \pd{\mu}P
	\ee
from which one can construct the second rank tensor on control space
	\be\label{eq:h}
	2\,\tr(A\otimes A^*)=
    (g_{\mu\nu}(\phi) -i\,\omega_{\mu\nu}(\phi)) d\phi^\mu\otimes d\phi^\nu,
	\ee
where $\otimes$ is the product of forms. The symmetric part is the natural notion of infinitesimal distance for projections, the Fubini-Study metric \cite{Nakahara},
	\be\label{eq:g}
	g_{\mu\nu}=\tr\,P_\perp \{ \pd{\nu} P,\,\pd{\mu}P\big\}
    =   \tr\, ( \pd{\nu} P)\,(\pd{\mu}P).
	\ee
The antisymmetric part gives  the adiabatic (or Berry's) curvature \cite{berry},
	\be\label{eq:omega}
	\omega_{\mu\nu}=i\,\tr \bigl(P_\perp [\pd{\mu}P,\pd{\nu}P]\bigr)
    =-i\,\tr \bigl(P [\pd{\mu}P,\pd{\nu}P]\bigr).
	\ee
The basic properties of $g$ and $\omega$ are direct consequences of the
Eqs.~(\ref{eq:g},~\ref{eq:omega}) and are summarized in the following statement.

    \begin{prop}\label{symmetries}
    $g\ge 0$ defines a metric on the space of controls $\cal M$. It is even under time-reversal and it is also even under electron-hole exchange, $g(P)=g(P_\perp )$.
 $\omega$ is a closed 2-form, $d\omega=0$, endowing $\cal M$ with a symplectic structure, if non-degenerate. It is anti-symmetric under time reversal and hence $\omega=0$ if $H(\phi)$ is time reversal invariant. $\omega$ is anti-symmetric under electron-hole exchange, $\omega(P)=-\omega(P_\perp)$.
    \end{prop}

\goodbreak
{\em Geometry of quantum response:}
\medskip

 For the sake of simplicity we consider the case where $H$ is finite dimensional matrix and $\Gamma_\alpha(H)$ are
real functions of $H$.  The (instantaneous) eigenvalues of $H$ shall be denoted by $\e_j$, and the corresponding eigenstates by $|j\rangle$.   The ground state shall be denoted by $j=0$.  The eigenvalues of $\lind$ (corresponding to eigenstates $|j\rangle\langle k|$)
shall be denoted by $\lambda_{jk}$. By inspection $\im(\lambda_{jk})= -\e_j+\e_k$ while $\re(\lambda_{jk})\le 0$, i.e. all eigenvalues lie in the (closed) left half-plane. If two states are degenerate $\e_j=\e_k$ then clearly $\lambda_{jk}=0$.

 It will be  convenient to introduce the notion of dimensionless dephasing rates $\gamma_{jk}\ge 0$
 associated with the pair of non-degenerate eigenstates
    \be\label{gammajk}
    \gamma_{jk}=-\frac{\re(\lambda_{jk})}{|\e_j-\e_k|}\ge 0.
    \ee
For  a pair of degenerate states, where both the numerator and denominator  vanish, the dephasing rate is
defined to be zero.

 We are now at a position where we can state our first main result.  In Appendix B we describe a formula relating the response matrix $f$ to  the spectral projections and their gradients for general dephasing Lindbladians.  The formula admits a particularly simple geometric interpretation provided
we make a specific choice for $\Gamma(H)$.  We shall first state the result in this special case and then comment on the general case.

    \begin{thm}\label{theorem}
 The matrix  of transport coefficients $f$, Eq.~(\ref{LinResp}),
associated to the adiabatic evolution 
characterized by a single dephasing  rate,  Eq.~(\ref{gammajk}),
$\gamma=\gamma_{j0}\ge 0$ to all other energy levels $j$, is given by
	\be\label{eq:omegaL}
	f=\frac \gamma {1+\gamma^2}\,g + \frac 1 {1+\gamma^2}\, \omega.
	\ee
 $g$ is the Fubini-Study metric of the ground state bundle  
 and $\omega$ its adiabatic curvature.
    \end{thm}

Let us make the following observations:
\begin{enumerate}
	\item The condition of a single dephasing rate is automatically satisfied in any two level system. It is also satisfied if $H$ has two 	degenerate eigenvalues.  When $H$ has $3$ or more distinct eigenvalues the condition is non-trivial and can be interpreted as a condition on the functional form of $\Gamma(H)$, e.g. $\Gamma(H)=\sqrt{\gamma(H-\e_0)}$ with $\e_0$ the lowest eigenvalue of $H$.
	\item The dissipative response is associated with the symmetric part of $f$ and is fully determined by the metric $g$. It vanishes in the limit $\gamma=0$ as it must---there is no dissipation in adiabatic unitary evolution with a gap condition. In contrast, in open dephasing systems, a gap condition does not provide protection from dissipation.
	\item The non-dissipative response is associated with the anti-symmetric part of $f$ and is fully determined by the adiabatic curvature and the dephasing rate.
 (An extension of
the Berry phase to systems with decoherence in the form of a complex phase is studied in \cite{gefen}.) 
	\item The theorem gives a geometric interpretation to both the dissipative and non-dissipative parts of the adiabatic quantum response.
	\item For weakly dephasing systems, $0\le \gamma\ll 1$, the dissipative response  depends linearly on $\gamma$ while the non-dissipative response depends quadratically on $\gamma$. Dephasing affects both the dissipative and non-dissipative response coefficients.
	\item Chern numbers are integers obtained by integrating the adiabatic curvature over a closed 2-dimensional control space.  The theorem says  that the relation between the control space average of transport coefficients and Chern numbers is a function of  the dephasing $\gamma$. Hence, topological quantum numbers are not robust against dephasing.
	  \item
    The theorem generalizes to the multi-dephasing rate case, where $\gamma_{j0}$ of Eq.~(\ref{gammajk}) is $j$-dependent, at the price of replacing $P_\perp\pd{\mu}P$ by a weighted sum of $P_j \pd{\mu}P$. The Fubini-Study metric is then replaced by a metric that does not have a standard name.
    \item The linear response formula, Eq.~(\ref{LinResp}), gives the leading
term in the adiabatic expansion of the response provided the spectrum of
$H(\phi)$ is independent of $\phi$. In the general case, when the eigenvalues
are $\phi$ dependent, the expansion Eq.~(\ref{LinResp}) has additional terms at low orders: One
which is not small, but depends on $\phi(t)$ only, and one which is
of the same order as $\dot\phi(t)$, but is given by a quadratic
expression in $\dot\phi(t')$ integrated over $t'\le t$. However, as these terms are not proportional to the instantaneous driving $\dot\phi(t)$ we shall not consider them here (cf. \cite{BR93}).

\end{enumerate}


\goodbreak
\medskip
{\em  Control space with compatible metric and symplectic structures: }
\medskip

Our second main result concerns the special class of Hamiltonians when the
metric $g$ and the symplectic structure $\omega$ are {\em compatible}. In the
case of a single pair of controls this is expressed by $\det g=\det
\omega$. In general and in terms of a basis in which both $g$ and $\omega$ are
$2\times2$-block diagonal, the two structures are compatible if they are so
inside each block. Equivalently, we say  that $g$ and $\omega$ are compatible if
    \be\label{compatible}
    \omega^{-1} g + g^{-1} \omega = 0.
    \ee
Eq.~(\ref{compatible}) implies the equality of the determinants.

By Proposition~\ref{symmetries}, compatibility is possible  {\em only when time reversal is broken}.  Below, we shall give three natural physical examples which give rise to compatible metric and symplectic structure. The three examples correspond to the three prototypical control  spaces ${\cal M}$:  the sphere, the plane and the torus, respectively.

\begin{thm}\label{compatibility}
Suppose that $g$ and $\omega$ are compatible. Then the inverse of the matrix of quantum response of Theorem \ref{theorem}
has the form
     \be \label{immunity}
    f^{-1}= \gamma g^{-1}+\omega^{-1}.
    \ee
   \end{thm}
The claim is easily verified by multiplying Eq.~(\ref{eq:omegaL}) by Eq.~(\ref{immunity}) and using Eq.~(\ref{compatible}).

Remarkably, the non-dissipative response associated with the antisymmetric part of $f^{-1}$  is {\em independent} of $\gamma$ and so immune
to dephasing.  It is determined by the adiabatic curvature alone, just like in the case of unitary evolution.

\medskip
{\em Compatibility test:}
\medskip

Given $P$ one is interested in simple tests that tell whether $g$ and $\omega$ are compatible without explicitly computing the matrices $g$ and $\omega$.

 \begin{thm}\label{compa}
 The following are equivalent.
 \begin{enumerate}
 \item[(i)] $g,\,\omega$ defined on $\cal M$ are compatible in the sense of Eq.~(\ref{compatible}).
 \item[(ii)] $\cal M$ has local holomorphic and antiholomorphic coordinates $z^j,\,\bar{z}^j$ making it into a complex manifold and the map $P(\phi)$
satisfies $P_\perp \bar{\partial}_j P = 0$.
\item[(iii)] The image of the map $P:{\cal M}\rightarrow \Gr(n+1,r;\mathbb{C})$ is a complex submanifold of the complex Grassmannian manifold $\Gr(n+1,r;\mathbb{C})$.
(Here $r=\rank(P)$ and $n+1$ is the dimension of the Hilbert space.)
 \end{enumerate}
Moreover when $P$ is a one dimensional projection  (i.e. $r=1$)
then (ii) is equivalent to the claim that $P$ may be expressed as $P={\ket{\psi}\bra{\psi}\over\braket{\psi}{\psi}}$
where $\ket{\psi}=\ket{\psi}_z$ is holomorphic, i.e. $\bar{\partial}_j\ket{\psi}=0$.
 \end{thm}


We show how the test (ii) implies compatibility (i) when $\cal{M}$ is
two dimensional. We write locally $z = \phi^1 + \tau \phi^2$,
where $\tau = \tau_1 + i \tau_2$ with $\tau_2 \neq 0$. Then
$ \bar{\partial} ={1\over2i\tau_2}( \tau \pd{1} - \pd{2})$ and
the assertion (ii) is equivalent to the assertion that
 \be\label{compatible1}
\tau A_1- A_2=0.
\ee
 Substituting this identity in Eq.~(\ref{eq:h}) gives
	\be
	g_{22}=|\tau|^2g_{11}, \quad g_{12}=\tau_1 g_{11}, \quad \omega_{12}=\tau_2 g_{11}.
	\ee
The equality of the determinants now follows by inspection.

For $P = \ket{0}\bra{0}$, (ii) is equivalent to $\pd{\bar{z}}\ket{0} \propto \ket{0}$ and hence
the possible $\bar{z}$ dependence of $\ket{0}_z$ is merely through a normalization/phase factor.

\medskip
{\em Complex structure--K\"ahler manifolds:}
\medskip


The compatibility condition is equivalent to the statement that
the operator $J=\omega^{-1}g$ acting on $T_\phi{\cal M}$, the tangent space at
$\phi$, satisfies $J^2=-1$. It can therefore be used to turn the real vector
space $T_\phi{\cal M}$ into a complex vector space by defining multiplication
by the scalar $\alpha+i\beta\in\mathbb{C}$ to be given by the action of  the
the operator $\alpha+\beta J$.

The compatibility condition Eq.~(\ref{compatible}) automatically holds for the Fubini-Study metric and the adiabatic curvature when the control space is the whole projective space
$\mathbb{PC}^n$ and the projections $P$ parametrize themselves. This is related to the fact that $\mathbb{PC}^n$ is a K\"ahler manifold. The same also holds for the Grassmanian manifold.
Our control space ${\cal M}$ may be viewed (through the map $\phi\mapsto P(\phi)$) as immersed in $\mathbb{PC}^n$ when $\rank(P)=1$ or in the Grassmanian
when $\rank(P)>1$. Indeed  we defined $g^{\cal M}$ and $\omega^{\cal M}$ on  ${\cal M}$ using this identification by restricting $g^{\mathbb{PC}^n},\omega^{\mathbb{PC}^n}$
to its immersion. 
$g^{\cal M},\omega^{\cal M}$ inherit the compatibility from $g^{\mathbb{PC}^n},\omega^{\mathbb{PC}^n}$
if and only if $\cal M$ is a {\em complex} submanifold of ${\mathbb{PC}^n}$
(or of $\Gr(n+1,r)$), in which case $\cal M$ inherits the K\"ahler property,
too. This leads to the Theorem~\ref{compa}.


We conclude by giving three examples of natural controlled physical systems whose ground state bundle has a metric compatible  with the curvature.

  \medskip
    {\em The Qubit family:}
The controlled Hamiltonian of a qubit is
	\be
	H= \hat \phi\cdot \vec{\sigma}, \quad \hat{\phi} \in \mathbb{S}^2.
	\ee
The control space ${\cal M}=\mathbb{S}^2$ is  the unit sphere.
The ground state projections are given by
	\be
	2P= {1-\hat\phi \cdot \vec{\sigma}}.
	\ee
The associated Fubini-Study metric and the symplectic form are readily computed from Eqs.~(\ref{eq:g},~\ref{eq:omega})
    \be
    2g_{ij} =  \pd i \hat\phi \cdot \pd j \hat\phi, \qquad
    2\omega_{ij} = - \hat\phi \cdot \pd i \hat\phi \times \pd j \hat\phi.
    \ee
Both give consistent areas (half the standard area) on $\mathbb{S}^2=\cal M$.

\medskip
    {\em Coherent states:}
The  Hamiltonian of a phase-space controlled oscillator is
	\be
	H(\zeta, \mu)=
    \frac 1 {2} (p-\mu)^2+\frac {1} 2\, (x-\zeta)^2
	\ee
where $(\zeta,\mu)\in \mathbb{R}^2=\cal M$, the Euclidean plane. The manifold of ground states is the coherent states. The controls are boosts and shifts hence
    \be
    A_1=P_\perp \partial_\zeta P=-iP_\perp [p, P]=-iP_\perp (p-\mu) P.
    \ee
Similarly,
    \be
    A_2=P_\perp \partial_\mu P=iP_\perp [x, P]=iP_\perp( x-\zeta) P.
    \ee
Since $a=(x-\zeta)+i(p-\mu)$  annihilates the ground state of the shifted oscillator, Eq.~(\ref{compatible1}) holds with $\tau=i $
    \be\label{acho}
    P_\perp a P=0.
    \ee

    \begin{figure}[htb]
    \centering
    \includegraphics[height=3 cm]{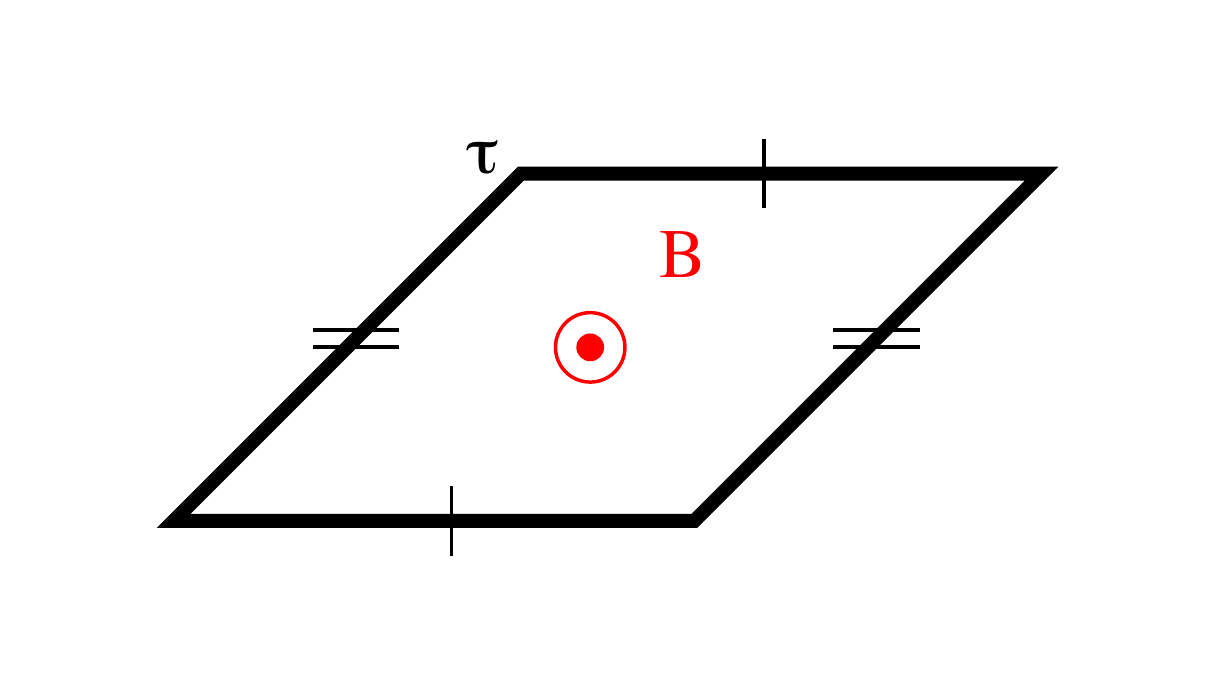}
    \caption{ A flat torus with skewness $\tau$ is obtained by identifying the parallel ends of the skewed parallelogram. A constant magnetic field with an integer number, $B$, of flux quanta penetrates the torus.  }\label{fig:skewed-torus}
    \end{figure}

\medskip
{\em Landau Hamiltonian:}
As a third example of a control space with K\"ahler structure consider the Landau Hamiltonian associated with a torus of unit area and skewness $\tau=\tau_1+i\tau_2$, $\tau_2 > 0$; (see Fig. \ref{fig:skewed-torus}). The torus is threaded by Aharonov-Bohm
magnetic fluxes  $(\phi_1,\phi_2)$  (see Fig.~\ref{fig:torus}) and is penetrated by constant magnetic field $2\pi B$, $B$ an integer, (see Fig~\ref{fig:skewed-torus}). The corresponding Hamiltonian is \cite{asz}
    \begin{align}
    H(\tau, \phi) =
     \frac{1}{\tau_2^2}D^*D\label{hamiltonian}
    \end{align}
where
$D=i(\tau\partial_x-\partial_y)-2\pi\tau ( B y +\phi)$ and
$\phi =\phi_1 -\phi_2/\tau$.
We impose the usual magnetic translation boundary conditions \cite{Zak}:
    \be
    \psi(x,y)= \psi(x+1,y)= e^{2\pi i Bx}\,\psi(x,y+1).
    \ee
The (single particle) ground state is $B$-fold degenerate with energy $E = 0$ independent of $\tau$ and $\phi$.
For simplicity we consider the case of a single particle and $B=1$. The lowest Landau level is then related to theta functions \cite{mumford}:
    \be
    \psi(x,y)=
    {(2 /\tau_2)^{1/4}} e^{-\pi|\tau|^2(\im \phi)^2/\tau_2} \,
    \sum_{n=-\infty}^{\infty}
    e^{2\pi in x}\,e^{i\pi\tau (y+n+\phi)^2}.
    \ee
Since, except for the normalization,  $\psi$ depends only on the ``holomorphic coordinate'' $\phi$ and not on the ``anti-holomorphic''
coordinate $\bar{\phi}$ it follows from Theorem \ref{compa} that $g$ and $\omega$ are compatible. Alternatively, the analyticity of
$\psi$ is also seen from that of $D$; note that the boundary conditions are
independent of $\phi$.

\medskip
{\em Conclusions:}
\medskip

This work studies the response of driven open quantum system governed by adiabatically evolving Lindbladians. When the Lindblad operator is {\em dephasing}, coherence and phase information is degraded but the energy of the system is still conserved. In this case, we find that the response admits a geometric interpretation induced from the behavior of the instantaneous Hilbert space projections.

We focus on the response associated with observables that can be derived by
the principle of virtual work from the Hamiltonian of the system. 
For these observables we find that, in contrast with the case of unitary evolution, there is residual dissipation: A spectral gap does not protect against dissipation in open systems in general.

Our first main result, Theorem \ref{theorem}, concerns the special case of dephasing two-level systems and, more generally, certain dephasing systems characterized by a single rate $\gamma$. In this case, the response matrix admits a simple geometric interpretation:
The non-dissipative transport is proportional to the adiabatic curvature $\omega$ and the dissipative response to the Fubini-Study metric $g$. Since the proportionality factor relating non-dissipative transport to the adiabatic curvature is a function of the dephasing rate $\gamma$,  it follows that the robustness of Chern numbers {\em does not} translate to a robustness of the transport coefficients against dephasing.

Our second main result, Theorem \ref{compatibility}, is concerned with a geometric mechanism
which provides protection against dephasing for certain non-dissipative response coefficients.
The mechanism leading to such a protection is the {\em compatibility} of the adiabatic curvature with the metric. In this  case, the relation between transport coefficients and Chern numbers is {\em independent} of the dephasing rate.

Our third main result concerns test of compatibility. In particular, we show that when the instantaneous stationary state are holomorphic functions of the driving parameters, the Fubini-Study metric is compatible with the adiabatic curvature. 

We conclude with three fundamental examples of compatible physical Hamiltonians associated with the three types of control spaces. The first example of a compatible system is the qubit Hamiltonian whose control space is the sphere, the second example of a compatible system is the Harmonic oscillator whose control space is the plane and the third example is a model of the quantum Hall effect on a torus.

{\bf Acknowledgments.}
We thank Misha Reznikov for suggesting that the Hall resistance is better behaved than the Hall conductance. This work is supported by the ISF and the
fund for Promotion of research at the Technion. M.~F. was
partially supported by UNESCO fund.

\section{Appendix A}\label{B}

Here we describe an example illustrating how adiabatic dephasing Lindbladian, with slaved dephasing term, naturally arise from stochastic unitary evolutions.
The example is a stochastic variant of Berry's paradigm of the notion of adiabatic curvature, namely, a spin 1/2 in a magnetic field \cite{berry}.  The evolution equation is
	\be
	\dot\rho=
        -i[ \vec{B}\cdot \vec{\sigma},\rho]
	\label{berry}\ee
 where $\vec{B}\in \mathbb{R}^3$ is a time dependent magnetic field and $\vec{\sigma}$ the vector of Pauli matrices.  The case considered by Berry is when $\vec{B}$ changes its orientation adiabatically say with fixed magnitude. We want now to consider the stochastic version of this model where the {\em magnitude}  of $\vec{B}$ is a stochastic variable while its orientation is changing smoothly (adiabatically) in time. Formally, this corresponds to replacing  $B$ in the evolution equation by
 	\[	
	\vec{B}\to W_t \vec{B}_0
	\]
where $W_t$ is (scalar, biased) white noise.  The canonical interpretation of Eq.~(\ref{berry}) as a stochastic differential equation goes through the Ito calculus \cite{ito}. To do so, it is convenient to expresses white noise in terms of the corresponding Brownian motion
	\[
	db_t:=W_t dt.
	\]
The rules of Ito calculus say that  $d\rho$ has to be expanded to first order in $dt$ and to second order in $db$.
This gives the stochastic evolution equation
	\[
	d\rho =
  -i[ H_0,\rho]db-\half (db)^2 [ H_0,[ H_0,\rho]],
\qquad H_0=\vec{\sigma}\cdot \vec{B}_0
	\]
where $B_0$ is the smooth (non stochastic) function of time.  In particular,
it follows that the (noise average) state  $\rho_a=\mathbb{E}(\rho)$  satisfies the adiabatic Lindblad equation
	\[
	\dot{ \rho_a} =
       \lind(\rho_a)=-i\mu[ H_0,\rho_a]-\half D [H_0,[H_0,\rho_a]]
	\]
where $\mu$ is the bias of the white noise $ \mu=\mathbb{E}(W_t)$ and $D$ its variance $\mathbb{E}(W_tW_s)=D\delta(t-s)$.  If $D\neq 0$, this gives a dephasing evolution where the dephasing is slaved to the time dependence of the Hamiltonian.
(A general framework for deriving  Lindbladian for general stochastic evolutions is described e.g. in  \cite{ref:stochastic}.)


\section{Appendix B}\label{A}
Here we outline the proof of Theorem~\ref{theorem}
by evaluating the terms proportional to $\dot\phi$ in Eq.~(\ref{LinResp}).

Let $P_j$ denote the spectral projections for $H$.  Since $\lind(P_j)=0$, the spectral projections  are instantaneous stationary states.
Let $P=P_0$ denote the projection on the ground state.
We also denote $E_{jk}=\ket{j}\bra{k}$. This is an eigenvector of the Lindbladian with eigenvalue $\lambda_{jk}$ i.e.  $\lind(E_{jk})=\lambda_{jk}E_{jk}$.

By the adiabatic theorem the states adheres to the spectral projection, $\rho(t)=P(\phi(t))+O(\dot\phi)$. The first order correction $\delta\rho$ to the state
satisfies
\be\label{dr}\lind(\delta\rho)=\dot{P},\ee
as can be seen from the substitution $\rho=P+\delta\rho$
into the Lindblad Eq.~(\ref{Lind}) and using $\lind(P)=0$.
The correction can be decomposed as $\delta\rho=\delta_\perp \rho + \delta_\parallel \rho$ into parts $\delta_\perp \rho\in\Ran \lind$ and $\delta_\parallel \rho\in\Ker \lind$, which are orthogonal with respect to the inner product defined by the trace. Note that  $\lind$ considered as a map on $\Ran \lind$  is invertible and that  $\dot P \in \Ran \lind$. Thus
Eq.~(\ref{dr}) implies $\delta_\perp\rho= \lind^{-1}(\dot P)$, where the inverse $\lind^{-1}$ is well defined.
In fact, since the eigenstates of $\lind$ are $E_{jk}$, one may readily write
\be\label{rp} \delta_\perp \rho=\lind^{-1}(\dot P)=\sum_{j\neq k}
\frac{ \bra{j}\dot P\ket{k}}{\lambda_{jk}}E_{jk}.\ee
Strictly speaking we restricted here to the case of simple eigenvalues;
more generally
$\bra{j}\dot P\ket{k}=0$ between degenerate eigenstates, whence
the appropriate reading of the sum~(\ref{rp}) is by omitting such pairs.

The complementary part $\delta_\parallel \rho$ may be determined as well (cf. \cite{BR93}) and happens to depend on history, but will not be needed.

Now, $\rho$ carries two contributions to the response Eq.~(\ref{LinResp}), of which the leading one, $\tr(P F_\mu)$ equals
$\partial_\mu\tr(PH)=\partial_\mu\e_0$. This term is not propotional to $\dot \phi$ and does not concern us (note that
it vanishes when the spectrum is independent of $\phi$, c.f. observation~8 after the theorem).
 As for the first order correction $\delta\rho$, two contributions
arise in turn through $F_\mu={\partial H\over\partial\phi^\mu}=\partial_\mu{\sum\e_j P_j}$. The first, $\sum (\partial_\mu\e_j)  P_j$, lies in $\Ker \lind$ and matches $\delta_\parallel \rho$.
This term does not concern us either and again vanishes when $\partial_\mu\e_j$ does.
The other part $\sum \e_j \partial_\mu P_j$ lies in $\Ran \lind$
and gives the requisite linear response term of the expectation value $\langle F_\mu\rangle= \tr(\rho\, \partial_\mu H)$:
 \be
    \sum_{i} \e_i \tr \bigl((\partial_\mu P_i) \, \delta_\perp\rho\bigr)
    =\sum_{i\neq 0} (\e_i-\e_0) \tr \bigl((\partial_\mu P_i) \, \delta_\perp\rho\bigr),
    \label{lin-res}
    \ee
where we used $\partial_\mu\sum_i P_i=0$. Eq.~(\ref{lin-res}) can now be written as $\sum_{i}(\e_i-\e_0) \mathcal{A}_i$, where
    \be
    \mathcal{A}_i=\sum_{j\neq k}
 \bra{k} \partial_\mu P_i\ket{j}
\frac{ \bra{j}\dot P\ket{k}}{\lambda_{jk}}.
    \label{lin-res2}
    \ee
Using
    \be
    \bra{j}\dot P\ket{k}=\bra{j}P_j\dot PP_k\ket{k}=\left(\delta_{k,0}+\delta_{j,0}\right)\bra{j}\dot P\ket{k}
    \label{Aa}
    \ee
it follows that the double sum in Eq. (\ref{lin-res2}) reduces to the single sum
    \be
    \mathcal{A}_i=\sum_{j\neq 0} \Bigl(\frac{\bra{0}\partial_\mu P_i\ket{j}\,\bra{j}\dot P\ket{0}}{\lambda_{j0}} + c.c. \Bigr).
    \label{lin-res3}
    \ee
Since $\overline{\lambda_{kj}}=\lambda_{jk}$ we get
 $\mathcal{A}_i$ is manifestly real as it must be.
Using the fact that (recall that $i\neq 0$)
        \be
        \bra{0}\partial_\mu P_i\ket{j}
    =\bra{0}P\partial_\mu P_i\ket{j}=-\bra{0}(\partial_\mu P) P_i\ket{j}=-\delta_{ij} \bra{0}\partial_\mu P \ket{j}
        \ee
we finally find
    \begin{align}
    \sum_{j\neq 0} (\e_j-\e_0) \mathcal{A}_j&=-\sum_{j\neq 0} \ {\frac {\e_j-\e_0} {\lambda_{j0}}}\, \tr\bigl(P (\partial_\mu P) P_j \dot P\bigr) + c.c. \nonumber\\
&=\sum_{j\neq 0} \ \frac 1 {i+\gamma_{j0}}\, \tr\bigl((\partial_\mu P) P_j \dot P\bigr) + c.c.,
    \label{lin-res4}
    \end{align}
where $\gamma_{j0}\ge 0$ is the dimensionless characterization of the spectral data of Eq.~(\ref{gammajk}).

Simplification  occurs for  $\gamma_{j0}$ is independent of $j$. This is, of course,
automatically the case for a two level system where $j$ takes one value $j=1$. (Similar simplification occurs when there is one dominant $1/\gamma_{j0}$.)    The sum over $j$ can now be carried out explicitly

    \be
    \langle F_\mu\rangle=
    \sum (\e_j-\e_0)\mathcal{A}_j=
    \frac {\gamma-i}{1+\gamma^2}  \,\tr \big((\partial_\mu P)P_\perp \dot P\big)
     + c.c.
    \label{lin-res-Final}
    \ee
Writing $\dot P= \sum_\nu (\partial_\nu P)\, \dot \phi$ we obtain the expression in the theorem.


\end{document}